\PassOptionsToPackage{table}{xcolor}
\documentclass[conference,a4paper]{APSIPA2025}
\usepackage{amsmath}
\usepackage{graphicx}
\usepackage{multirow}
\usepackage{pifont} 
\usepackage{makecell}
\usepackage{booktabs}
\usepackage{multirow}
\usepackage{threeparttable}
\usepackage{amsmath}          % for align environment, \text, etc.
\usepackage{algorithm}        % for floating algorithm environment
\usepackage{algpseudocode}    % for algorithmic pseudocode (\begin{algorithmic}…)
\usepackage{url} 
\usepackage{comment}
\usepackage{amsmath, amssymb}

\usepackage{geometry}
\usepackage{fancyhdr}
\usepackage{todonotes}

\fancypagestyle{firststyle}{
  \fancyhf{}
  \fancyhead[C]{2025 Asia Pacific Signal and Information Processing Association Annual Summit and Conference (APSIPA ASC)}
}

\begin{document}

\title{SegReConcat: A Data Augmentation Method for Voice Anonymization Attack}

\author{
\authorblockN{
Ridwan Arefeen\authorrefmark{1},
Xiaoxiao Miao\authorrefmark{2},
Rong Tong \authorrefmark{1},
Aik Beng Ng\authorrefmark{3},
Simon See\authorrefmark{3}
}

\authorblockA{
\authorrefmark{1}
Singapore Institute of Technology \\
E-mail: 2403754@sit.singaporetech.edu.sg, tong.rong@singaporetech.edu.sg}

\authorblockA{
\authorrefmark{2}
Duke Kunshan University, China \\
E-mail: xiaoxiao.miao@dukekunshan.edu.cn
}

\authorblockA{
\authorrefmark{3}
NVIDIA AI Technology Centre \\
E-mail: \{aikbengn,ssee\}@nvidia.com  }
}

\maketitle
\thispagestyle{firststyle}
\pagestyle{fancy}

\begin{abstract}
  Anonymization of voice seeks to conceal the identity of the speaker while maintaining the utility of speech data. However, residual speaker cues often persist, which pose privacy risks. We propose \textbf{SegReConcat}, a data augmentation method for attacker-side enhancement of automatic speaker verification systems. SegReConcat segments anonymized speech at the word level, rearranges segments using random or similarity-based strategies to disrupt long-term contextual cues, and concatenates them with the original utterance, allowing an attacker to learn source speaker traits from multiple perspectives. The proposed method has been evaluated in the VoicePrivacy Attacker Challenge 2024 framework across seven anonymization systems, SegReConcat improves de-anonymization on five out of seven systems\footnote{Code can be found at \url{https://github.com/monkeyDarefeen/SegReConcat_augmenter}.}. 
\end{abstract}

\section{Introduction}
\label{sec:intro}
%what is the task, why it is important, how other people do this task, what are the challenges/drawbacks/unexplored aspects, what we proposed and the contributions  (1-1.5 pages)
Voice data inherently contains rich personal information, including speaker identity, emotional state, and potential medical conditions. When raw speech is shared, for example, with cloud services or on social media, adversaries can exploit these biometric cues to conduct identity theft or enable targeted advertising \cite{GDPR}. As a result, voice privacy has become a critical research objective, often viewed as an ongoing game between users (defenders) and attackers (adversaries). 

On one hand, users employ voice anonymization or deidentification techniques to modify original speech utterances to conceal the speaker's identity while preserving linguistic content and permissible paralinguistic attributes. This process generates anonymized speech that can be safely shared. The VoicePrivacy Challenges (VPC) have played a key role in advancing this field by providing benchmarks \cite{Tomashenko_2020,tomashenko2022voiceprivacy2022challengeevaluation, 10603395,tomashenko2024voiceprivacy} and fostering the development of anonymization technologies \cite{champion2023, panariello2024speakeranonymizationusingneural,meyer2023prosody, shamsabadi2022differentially, mawalim2022speaker, yao2024musa, miao22_odyssey, miao2023language, yao2024distinctive, xinyuan2024hltcoe, cai2024privacy, ghosh2024anonymising, zhang2023voicepm, miao2024benchmark, miao2025adapting, yao2025easy, das2025}.

On the other hand, attackers leverage their knowledge of anonymization systems to infer residual identifying information from anonymized speech. While most research focuses on improving anonymization from the users' perspective and assumes fixed attack scenarios, where attackers primarily rely on automatic speaker verification (ASV) systems such as ECAPA-TDNN \cite{Desplanques_2020} to trace source speaker information, relatively little attention has been paid to developing comprehensive methods for evaluating the robustness of anonymization against determined adversaries. 
To facilitate research in this area, the same committee that organized the VPCs launched the first VoicePrivacy Attacker Challenge (VPAC) in 2024 \cite{tomashenko2024voiceprivacyattackerchallengeevaluation, Tomashenko_2025}. The challenge provided anonymized datasets generated by various anonymization systems \cite{tomashenko2024voiceprivacy}, along with a baseline attacker system. Participants developed ASV systems to evaluate the robustness of different anonymization methods.
The top submissions primarily improved upon three aspects:
(i) Data augmentation: the SpecWav attack \cite{li2025specwav}, combining wav2vec2.0 features \cite{baevski2020wav2vec} with spectrogram resizing data augmentation technique.
(ii) Neural network-based embedding extraction: LoRA-adapted ResNet34 \cite{hu2022lora,he2016deep, lyu2025fast}, as well as fine-tuned the TitaNet-Large model \cite{mawalim2025fine}.
(iii) Backend models: \cite{zhang2025attacking} utilize ECAPA-TDNN feature extractor trained on mixed datasets with a PLDA-based \cite{prince2007probabilistic, ioffe2006probabilistic} scoring backend trained on anonymized data, incorporating SpecAugment \cite{Park_2019} for additional robustness.
In addition, other submissions explored alternative approaches, such as using different distance metrics and speaker embedding normalization techniques \cite{xinyuan2025hltcoe}. 

Although the attack strategies described above have proven effective in exposing weaknesses in anonymization systems \cite{Tomashenko_2025}, certain domains remain underexplored. Recent studies have shown that identity-revealing cues may persist in prosodic, phonetic, and linguistic content features that remain unchanged after anonymization \cite{gaznepoglu2025disentanglement, miao2025adapting, xinyuan2024hltcoe, cai2024privacy}, also in speech temporal dynamics \cite{tomashenko2025analysis, tomashenko2025exploiting}. For instance, both linguistic content and prosody can inadvertently preserve speaker-specific characteristics in the anonymized output \cite{panariello2024speakeranonymizationusingneural}. Moreover, inference attacks that match pseudo-speaker embeddings against a known speaker pool have been shown to outperform generic ASV models in de-anonymization tasks \cite{bauer2025inferenceattacksxvectorspeaker}.

Motivated by these findings, this paper follows the VPAC 2024 framework and proposes a novel attacker-side data augmentation technique, called \textbf{SegReConcat}, where an anonymized speech utterance is \textbf{seg}mented into word segments, then \textbf{re}arranges the words according to different strategies, and finally \textbf{concat}enates with the input anonymized speech. The rearranged anonymized speech is generated by permuting the word sequence of the anonymized utterance\footnote{Note that the anonymized speech provided by VPAC 2024 always retains the original word sequence to preserve utility, such as content and emotion, which are important to users, while still ensuring strong concealment of the original speaker’s identity. However, as an attacker, one can reorder the sequence to exploit residual speaker identity information.}.

Since both content and prosodic information can reveal speaker identity potentially in both short-term and long-term contexts, an ASV model trained on anonymized speech with the original word order may inadvertently learn residual speaker traits from long-term sequential dependencies, which are easier to capture. By disrupting the word order, \textbf{SegReConcat} breaks these long-term contextual cues, thereby forcing the model to focus on extracting residual speaker information from short-term, word-level features. This makes the attack more targeted and potentially more effective in revealing subtle identity cues. Additionally, by concatenating the rearranged speech with the original anonymized speech to create augmented inputs, the model is enabled to learn source speaker traits from multiple perspectives.

We explore different strategies for word rearrangement, including randomly shuffling words and grouping similar words based on similarity scores. We evaluated our attack method and demonstrated consistent reductions in equal error rate (EER) across 5 out of 7 anonymization systems provided by VPAC 2024. 

\section{Related Work}
% 1 page
In this section, we present the official design of the VPAC 2024, which provides the framework for this study. 
%This includes the definition of specific goals, an overview of seven anonymization models, and the objective evaluation metrics. 
We also review existing attacker approaches, with a focus on data augmentation techniques in this domain and their limitations.

\subsection{The VoicePrivacy Attacker Challenge}
\label{subsec:vp-challenge}

The VPAC 2024 \cite{tomashenko2024voiceprivacyattackerchallengeevaluation, Tomashenko_2025} offers a standardized framework to evaluate voice anonymization systems under a semi-informed attacker model. Attackers build ASV systems using anonymized speech to determine whether two utterances belong to the same speaker, aiming to lower EER, lower EER indicates a more effective recovery of speaker identity.

The attacker evaluates anonymized LibriSpeech \cite{panayotov2015librispeech} corpus for seven anonymization systems, each system applies a different anonymization strategy, such as disentanglement-based speaker embedding manipulation and codec-based synthesis. 
%The attacker evaluates seven anonymization systems, 
These systems include three VPC 2024 baselines (\textit{B3}–\textit{B5}) and four top submissions (\textit{T8-5}, \textit{T10-2}, \textit{T12-5}, \textit{T25-1}) \cite{tomashenko2024voiceprivacy}.

\begin{itemize}
  \item \textit{B3} \cite{b3}: phonetic transcription with pitch/energy modification and generated pseudo-speaker embeddings
  \item \textit{B4} \cite{panariello2024speakeranonymizationusingneural}: a neural audio codec with language modeling of semantic tokens
  \item \textit{B5} \cite{champion2023}: vector-quantized bottleneck (VQ-BN) features from an ASR model with original pitch.
  \item \textit{T8-5} \cite{xinyuan2024hltcoe}: a hybrid scheme randomly choosing between (1) ASR(Whisper)+TTS(VITS) and (2) k-NN voice conversion on WavLM features.
  \item \textit{T10-2} \cite{yao2024npu}: a neural codec (using VQ) that explicitly disentangles linguistic content, speaker identity, and emotion.
  \item \textit{T12-5} \cite{Kuzmin2024ntu}: an extension of B5 with additional pitch smoothing.
  \item \textit{T25-1} \cite{Gu2024ustc}: VQ-BN content features with global style tokens and emotion transfer for speaker style.
\end{itemize}
Each system's anonymized data includes both training and evaluation subsets. The training set LibriSpeech \textit{train-clean-360} contains approximately 104,014 anonymized utterances derived from 921 speakers. The development and test sets are anonymized versions of LibriSpeech \textit{dev-clean} and \textit{test-clean} respectively, each comprising both female-female and male-male speaker trial pairs. 
\begin{figure}[!t]
    \centering
\includegraphics[width=0.4\textwidth, trim=20pt 20pt 20pt 10pt, clip]{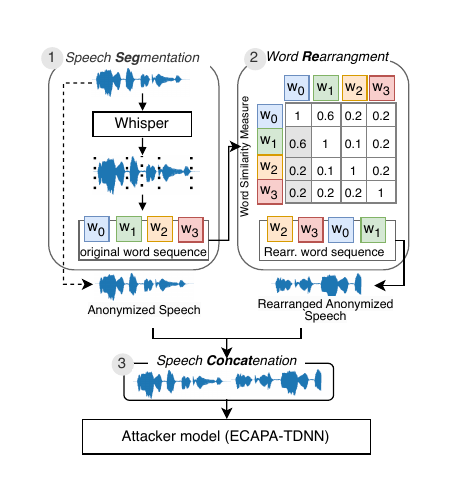}
    \caption{ (1) SegReConcat Flow Diagram}
    \label{fig:1}
\end{figure}

\subsection{Existing Attacker Approaches} 
% focus on the data augmentation approaches you will show in the experiments as the baselines, 1) rely on the other datasets, e.g. add noise/music/reverbaration 2) augmentation by changing/masking features, e.g. specaugment, resize the spectram.. 
Attacker systems have been developed from various perspectives, as outlined in Section \ref{sec:intro}. This section primarily reviews attacker-side data augmentation techniques, particularly those involving modifications to acoustic features that can be applied to many attacker systems. These include methods such as SpecAugment \cite{park2019specaugment}, which randomly masks time frames or frequency bands.
One system submitted to the VPAC 2024 employed SpecAugment alongside a mixed-data training strategy, combining anonymized and original speech to improve speaker discriminability.
Additionally, techniques that resize the spectrogram by stretching or compressing it along the frequency axis have been proposed for attacker models \cite{li2025specwav}. After stretching, the excess high-frequency components are trimmed; after compression, zero-padding is applied to the high-frequency band to maintain consistent frequency dimensions before and after resizing. 

The motivation behind these methods is to introduce perturbations into the spectrogram, forcing the attacker model to learn speaker-relevant features that are robust to such distortions. Unlike existing data augmentation techniques, this work proposes \textbf{SegReConcat}, a perceptually motivated data augmentation method. Details follow in the next section. %that permutes the word sequences to disrupt long-term dependencies in linguistic and prosodic patterns. By concatenating normal and reordered word sequences, \textbf{SegReConcat} encourages the attacker model to learn speaker information from varied perspectives of word sequence structure. The details of this method will be explained in the following section.

\section{Proposed SegReConcat Data Augmentation}
% assumptions and overall pipeline description (figure), followed by the details of each component, using mathematics symbols (1 page)
%\todo[inline]{TR:somehow,  figure 1 has low resolution to me, Xiaoxiao:updated}
This section details the proposed \textbf{SegReConcat} framework, which consists of three stages: \textbf{Seg}mentation, \textbf{Re}arrangement, and \textbf{Concat}enation, as shown in Figure~\ref{fig:1}. 
It takes an anonymized utterance, segments it at the word level, rearranges the words, and finally concatenates the rearranged sequence with the original audio.

The motivation for segmenting and rearranging the anonymized word sequence is to break the continuous content flow of speech and thereby disrupt the long-term temporal dynamics of speaker information. We assume that certain prosodic and phonetic patterns, such as coarticulation style, pitch dynamics, and spectral characteristics, remain speaker-dependent even after anonymization. By reordering word-level segments, we aim to amplify these residual stylistic cues in a way that is consistent for each speaker.

The motivation for combining rearranged speech with its original anonymized version within a single training instance, so the ASV model observes different realizations of the same linguistic content. This encourages learning speaker traits invariant across different segment orders, promoting reliance on speaker-specific acoustic patterns rather than content structure. Although data volume doubles, the structural correlation introduced enhances speaker-discriminative representation learning.

We now elaborate on each step of the proposed \textbf{SegReConcat} method.
%\todo[inline]{TR:re-organized the following, please check if missing anything}

\subsection{Segmentation}
\label{sec:seg}
Given an anonymized utterance $u$, the first step is to perform word-level segmentation using the Whisper-medium ASR model\footnote{\url{https://huggingface.co/openai/whisper-medium}}, which achieves a 2.9\% word error rate (WER) on LibriSpeech \textit{test-clean} \cite{radford2022robustspeechrecognitionlargescale}. This high accuracy on the LibriSpeech dataset ensures reliable word boundary detection and decompose the utterance into $N$ discrete word segments represented as $\boldsymbol{w}^o = {w_1, w_2, \dots, w_N}$.

\subsection{Rearrangement}
Next, we apply a rearrangement function \( g \) to the sequence \( \boldsymbol{w}^o \), a permuted word order \( \boldsymbol{w}^r = g(\boldsymbol{w}^o) \). The aim is to disrupt the original content flow and long-range temporal dependencies that may preserve speaker characteristics, even after anonymization. The function \( g \) processes each utterance individually.
Three rearrangement strategies are introduced. One straightforward method is to randomly shuffle the original sequence \( \boldsymbol{w}^o \) to remove natural sequential cues. This is referred to as random rearrangement \textbf{(RR)}.

The other two strategies are similarity-based approaches, aiming to maximize the total similarity between consecutive words in \( \boldsymbol{w}^r \). This aims to investigate whether grouping similar words can amplify speaker style characteristics, thereby enhancing source speaker learning in ASV model. To measure word similarity, we consider both acoustic and semantic features.

\subsubsection{Acoustic Feature–Based Rearrangement \textbf{(AR)}}
Let us denote the MFCC feature matrices for two word segments \( w_i \) and \( w_j \) as \( \mathbf{M}_i \in \mathbb{R}^{d \times T_i} \) and \( \mathbf{M}_j \in \mathbb{R}^{d \times T_j} \), respectively, where \( d = 13 \) is the MFCC dimension, and \( T_i \) and \( T_j \) are the frame lengths of the corresponding word segments. Considering the different frame lengths across word segments, the Dynamic Time Warping (DTW) distance \cite{muller2007dynamic}, using Euclidean frame-wise cost, is computed to measure the acoustic similarity between segments, and is denoted as
$D_{ij} = \mathrm{DTW}\bigl(\mathbf{M}_i, \mathbf{M}_j\bigr)$, and is then converted into a similarity score via
$s_{ij} = \frac{1}{1 + D_{ij}}$.

To obtain a \( \boldsymbol{w}^r \) that optimizes the total similarity between consecutive words, we perform a greedy traversal: starting from the first segment (or an initial silence segment, if detected), we repeatedly append the unused segment with the highest similarity to the current one. Any remaining segments (if not reachable) are appended in arbitrary order.

\subsubsection{Semantic Feature–Based Rearrangement \textbf{(SR)}}
Besides using MFCC features to measure word similarity, we explore semantic feature measurements, as they are expected to carry more content information and better group words with similar pronunciations.
Each segment is passed through the encoder of the same pretrained Whisper-medium ASR model used for utterance segmentation in Section~\ref{sec:seg} to obtain a hidden segment-level representation \( \mathbf{h} \in \mathbb{R}^{2 \times d'} \), obtained by concatenating the mean and max pooled frame-level features over time, where \( d' \) is is the dimension of the last layer of the encoder, i.e. 1024.

Pairwise word segment $w_i$ and $w_j$ similarities are computed using cosine similarity: $
h_{ij} = \frac{\mathbf{h}_i^\top \mathbf{h}_j}{\|\mathbf{h}_i\| \, \|\mathbf{h}_j\|}.$
Similar to \textbf{AR}, $\boldsymbol{w}^r$ is constructed by selecting the segment with the highest average similarity as the starting point, and then iteratively appending the most similar remaining segment at each step.

\subsection{Concatenation}
%The final step is to concatenate the original and rearranged sequences:
To further enhance the diversity and robustness, we concatenate the original $\boldsymbol w^o$  and the rearranged sequence $\boldsymbol w^r$, forming an augmented sequence $\boldsymbol{w}^{\text{aug}} = \text{concat}(\boldsymbol{w}^o, \boldsymbol{w}^r)$, totaling $2N$ segments. 

\begin{table}[t]
\renewcommand{\arraystretch}{1.1}
\caption{The table reports averaged EER on T8-5 anonymization system.}
\label{tab:eer-t8-5}
\centering
\begin{tabular}{@{}l c@{}}
\toprule
\textbf{Methods} & \textbf{Avg EER (\%)} \\
\midrule
No Augmentation  & 37.39 \\
\midrule
SpecAug & 37.93 \\
\midrule

%Mix No Augmentation (50\%) + RR (50\%) & 31.74 \\
%Frame-Shuffle & 49.81 \\
%\hline
\textbf{RR} & 38.36 \\
\quad + Concatenation & 26.51 \\
\quad + \cellcolor{gray!7}{Concatenation + SpecAug} & \cellcolor{gray!7}{\textbf{25.94}} \\
\midrule

\textbf{AR} & 38.34 \\
\quad + Concatenation & 26.80 \\
\quad + \cellcolor{gray!7}{Concatenation + SpecAug} & \cellcolor{gray!7}{27.82} \\
\midrule

\textbf{SR} & 39.10 \\
\quad + Concatenation & 27.42 \\
\quad + \cellcolor{gray!7}{Concatenation + SpecAug} &
\cellcolor{gray!7}{32.70} \\
\bottomrule
\end{tabular}
\vspace{-0.5em}
\end{table}

%Currently baseline is BL-1024, 
%need to realize if I should use BL-1024 or SpecAug as baseline***
\begin{table*}[ht!]
\caption{Performance comparison of the ECAPA-TDNN attacker systems with and without SegReConcat (using \textbf{RR}), in terms of EER (\%), against seven anonymization systems.}
\vspace{-4mm}
\begin{center}
\resizebox{0.95\textwidth}{!}{
\begin{tabular}{c c c c c c c c c c c}
  \toprule
  \multirow{2.5}{*}{\makecell{Anonymization System}} & \multirow{2.5}{*}{\makecell{Attacker System}} & \multirow{2.5}{*}{SpecAugment} & \multicolumn{3}{c}{EER on dev-clean subset} & \multicolumn{3}{c}{EER on test-clean subset} & \multirow{2.5}{*}{Avg. EER} \\
  \cmidrule(lr){4-6} \cmidrule(lr){7-9}
  & & & Female & Male & Avg. & Female & Male & Avg. & \\
  \midrule
  \multirow{2}{*}{B3} 
  & Baseline & \ding{55} & \textbf{27.56} & 22.98 & \textbf{25.27} & 29.01 & 27.84 & 28.43 & 26.85 \\
  & \textbf{SegReConcat+SpecAug} & \ding{51} & 27.58 & \textbf{22.98} & 25.28 & \textbf{28.79} & \textbf{24.28} & \textbf{26.53} & \textbf{25.91} \\
  \midrule
  \multirow{2}{*}{B4} 
  & Baseline & \ding{55} & 35.09 & 28.73 & 31.91 & 26.10 & 29.62 & 27.86 & 29.89 \\
  & \textbf{SegReConcat} & \ding{55} & \textbf{30.67} & \textbf{24.99} & \textbf{27.83} & \textbf{23.54} & \textbf{25.17} & \textbf{24.36} & \textbf{26.09} \\
  \midrule
  \multirow{2}{*}{B5} 
  & Baseline & \ding{55} & \textbf{32.53} & \textbf{29.21} & \textbf{30.87} & \textbf{30.66} & \textbf{31.21} & \textbf{30.94} & \textbf{30.90} \\
  & \textbf{SegReConcat+SpecAug} & \ding{51} & {35.62} & {30.24} & {32.93} & {33.25} & {32.52} & {32.89} & {32.91} \\
  \midrule
  \multirow{2}{*}{T8-5} 
  & Baseline & \ding{55} & 33.49 & 38.81 & 36.15 & 39.58 & 37.67 & 38.63 & 37.39 \\
  & \textbf{SegReConcat+SpecAug} & \ding{51} & \textbf{26.42} & \textbf{26.86} & \textbf{26.64} & \textbf{25.55} & \textbf{24.94} & \textbf{25.25} & \textbf{25.94} \\
  \midrule
  \multirow{2}{*}{T10-2} 
  & Baseline & \ding{55} & 45.74 & 40.84 & 43.29 & 39.60 & 41.43 & 40.52 & 41.90 \\
  & \textbf{SegReConcat+SpecAug} & \ding{51} & \textbf{41.62} & \textbf{35.25} & \textbf{38.44} & \textbf{37.59} & \textbf{39.20} & \textbf{38.40} & \textbf{38.41} \\
  \midrule
  \multirow{2}{*}{T12-5} 
  & Baseline & \ding{55} & \textbf{34.94} & \textbf{31.70} & \textbf{33.32} & \textbf{30.65} & \textbf{35.02} & \textbf{32.84} & \textbf{33.08} \\
  & \textbf{SegReConcat+SpecAug} & \ding{51} & {35.94} & {33.54} & {34.74} & {34.31} & {36.97} & {35.64} & {35.19} \\
  \midrule
  \multirow{2}{*}{T25-1} 
  & Baseline & \ding{55} & 40.77 & 37.60 & 39.19 & 39.60 & 38.53 & 39.06 & 39.12 \\
  & \textbf{SegReConcat} & \ding{55} & \textbf{38.35} & \textbf{35.10} & \textbf{36.73} & \textbf{37.77} & \textbf{35.41} & \textbf{36.59} & \textbf{36.66} \\
  \bottomrule
\end{tabular}    
}
\end{center}
\label{tab:system_comparison}
\vspace{-4mm}
\end{table*}

\section{Experiments}
%overall description, e.g. "To evaluate the effectiveness of the proposed method, we first take one anonymized system T8-5... to find the best configuration, we then apply the proposed strategies to the systems and verify the improvement...

To evaluate the effectiveness of the proposed \textbf{SegReConcat} method, we conducted systematic experiments within the framework of the VPAC 2024. %The primary goal is to assess whether our augmentation strategy can be used to enhance speaker verification (attacking) performance on the anonymized speech.
We begin by focusing on one anonymization system, \textit{T8-5}, selected for its relatively strong privacy performance and architectural diversity. This serves as our development benchmark for investigating various configurations of the proposed method, including different segment reordering strategies (e.g., random shuffle \textbf{RR}, acoustic-based \textbf{AR}, and semantic-based \textbf{SR}, with and without concatenation, and in combination with additional spectral augmentation.

\subsection{System Configurations}
\label{subsec:ds-sysconf}
%data
We fully follow the VPAC 2024 \cite{tomashenko2024voiceprivacyattackerchallengeevaluation} protocol as previously described in Section~\ref{subsec:vp-challenge} %which provides anonymized LibriSpeech \cite{panayotov2015librispeech} corpus for seven anonymization systems, each system applies a different anonymization strategy, including embedding manipulation, codec-based synthesis, and content-style disentanglement, as previously described in Section~\ref{subsec:vp-challenge}.
%Each system's anonymized data includes both training and evaluation subsets. The training set \textit{train-clean-360} contains approximately 104,014 anonymized utterances derived from 921 speakers. The development and test sets are anonymized versions of LibriSpeech \textit{dev-clean} and \textit{test-clean} respectively, each comprising both female-female and male-male speaker trial pairs. 
%system setting
and experiment with the proposed data augmentation method using different configurations on all seven anonymization systems. All experiments share the same speaker verification back-end, which is based on the ECAPA-TDNN architecture with 1,024 channels in the convolution frame layers and implemented via SpeechBrain\footnote{\url{https://speechbrain.readthedocs.io/}}.
%which has demonstrated strong performance in both standard and privacy-aware ASV tasks. 
Speaker embeddings are extracted from the final layer, and similarity scoring is performed using cosine distance between averaged enrollment and test embeddings. The model is trained from scratch using anonymized training data only. Regarding the training hyperparameters, we primarily follow the default settings in SpeechBrain, however, unlike the default of 10 epochs, we train the model until the development loss stops decreasing and select the model that achieves the best development loss.
EER is used as the primary metric for assessing re-identification performance. A lower EER indicates a stronger attack. 

\subsection{Experimental Results and Analysis}
%explain the experiments, 
\subsubsection{Ablation Study on T8-5}
%As described in Section~\ref{subsec:ds-sysconf}, Table~\ref{tab:eer-t8-5} represents the results of the data augmentation methods we have experimented. Each augmentation strategy is evaluated by training an ECAPA-TDNN\cite{Desplanques_2020} based ASV model on the anonymized speech. As shown in Table~\ref{tab:eer-t8-5}, the proposed method improves re-identification performance compared to several baselines, with the best result achieved when using word-level rearrangement, concatenation, and SpecAug, yielding an average EER of 25.94\%—a substantial improvement over the T8-5 baseline (40.76\%).
Table~\ref{tab:eer-t8-5} compares the EER results on \textit{T8-5} without data augmentation, with the commonly used SpecAug, along with our proposed \textbf{SegReConcat} under different configurations.
The first observation is that SpecAug achieves a very similar EER to the baseline system without any data augmentation. And using rearrangement data alone, regardless of which strategy, results in slightly higher EERs, indicating worse attack performance. One possible reason is that training solely on rearranged data confuses the ASV attacker system, preventing it from learning speaker-discriminative cues from the disordered sequences. However, applying the concatenation method on top of the RR, AR, or SR strategies reduces the EER by around 10\% (from 37.39\% to around 27\%), demonstrating an effective attacking ability. 

If we compare the results of \textbf{SegReConcat} using different rearrangement approaches, the trend is that `RR + concatenation` (26.51\%) performs slightly better than `AR + concatenation` (26.80\%), with `SR + concatenation` (27.42\%) being slightly worse, though overall the differences are minor. Notably, the similarity-based rearrangement approaches (AR/SR) require additional computation for similarity measurement yet fail to surpass the performance of simple random arrangement (RR). This suggests that RR's inherent sequence diversity may help prevent the ASV model from overfitting to artificially constructed acoustic patterns.

Next, considering that SpecAug masks the acoustic features while \textbf{SegReConcat} permutes the words, the two data augmentation methods are fundamentally different. It is interesting to investigate whether these augmentations are complementary. Therefore, we combine both methods (highlighted in grey) and find that the improvements are not consistent,  'RR + Concatenation + SpecAug` achieves the lowest EER (25.94\%).

In summary, through the ablation study on \textit{T8-5}, we have verified that the proposed \textbf{SegReConcat} is effective, with either the `RR + concatenation` or the `RR + concatenation + SpecAug` configuration performing the best. We now proceed to experiments on the other systems using these configurations.

\subsubsection{Results Across Seven Anonymized Systems}

Experimental results on all seven anonymization systems are shown in Table \ref{tab:system_comparison}, where \textbf{SegReConcat} specifically denotes the use of RR as the similarity metric during the rearrangement stage. 
We evaluated both \textbf{SegReConcat} alone and in combination with SpecAug for all anonymization systems, and report the best results achieved. 
Among the seven systems, five show lower EERs and thus stronger attacks. Of these, two achieve the lowest EER using \textbf{SegReConcat} alone, while the other three achieve the lowest EER using \textbf{SegReConcat} in combination with SpecAug. The absolute average EER reductions range from 1\% to 3\%, with the exception of \textit{T8-5}, which achieves an average absolute EER reduction of around 11\%.

For the two systems that did not show a decrease in EER after applying the proposed data augmentation, i.e., \textit{B5} and \textit{T12-5} (see system descriptions in Section~\ref{subsec:vp-challenge}), it is noteworthy that both use VQ layers to quantize the continuous self-supervised learning features. A possible explanation is that our proposed data augmentation method is designed to disrupt the continuous content flow of speech and interfere with the long-term temporal dynamics of speaker information. However, the VQ process already discretizes the SSL features and removes continuity, potentially rendering \textbf{SegReConcat} less effective. Note that \textit{T25-1} also uses VQ-BN features, but \textbf{SegReConcat} still achieves a lower EER. This may be due to its additional use of emotion transfer technology, which could leak temporal speaker dynamics. These dynamics can be further disrupted by \textbf{SegReConcat}, enhancing the exploitation of speaker-specific features for the attacker's ASV model.

\section{Conclusions}
This work investigates the vulnerability of the current voice anonymization systems from the attacker’s perspective by introducing a novel augmentation strategy, \textbf{SegReConcat}, which can be applied to any back-end ASV model to enhance re-identification.
%designed to enhance ASV-based re-identification. 
This study is motivated by the observation that anonymized speech often retains residual speaker-specific cues in prosody and articulation. The proposed method leverages segment-level reorganization and concatenation to amplify such cues and improve speaker discriminability.
Experimental results on the \textit{T8-5} anonymization system demonstrate that the proposed approach yields improvement in speaker verification accuracy, achieving an 11\% absolute reduction in average EER compared to the baseline. Further evaluation across all seven anonymization systems in the VoicePrivacy 2024 Attacker Challenge\cite{tomashenko2024voiceprivacyattackerchallengeevaluation}  shows superior attacking performance for five systems, confirming the robustness and generalizability of the method.

These findings highlight limitations in current anonymization pipelines, particularly in their inability to fully suppress subtle speaker identity traces. Our results emphasize the need for future anonymization systems to consider attacker-informed augmentations and prosodic invariance more explicitly in their design.

\section{Acknowledgment}
This work is supported by Ministry of Education, Singapore, under its Academic Research Tier 1 (R-R13-A405-0005). Xiaoxiao Miao is the corresponding author and this work was conducted while she was at SIT. Thanks Timothy Liu from Nvidia AI Technology Centre for his insightful  suggestions.
%\printbibliography

\bibliographystyle{IEEEtran}
\bibliography{main}

% Generated by IEEEtran.bst, version: 1.14 (2015/08/26)
\begin{thebibliography}{10}
\providecommand{\url}[1]{#1}
\csname url@samestyle\endcsname
\providecommand{\newblock}{\relax}
\providecommand{\bibinfo}[2]{#2}
\providecommand{\BIBentrySTDinterwordspacing}{\spaceskip=0pt\relax}
\providecommand{\BIBentryALTinterwordstretchfactor}{4}
\providecommand{\BIBentryALTinterwordspacing}{\spaceskip=\fontdimen2\font plus
\BIBentryALTinterwordstretchfactor\fontdimen3\font minus \fontdimen4\font\relax}
\providecommand{\BIBforeignlanguage}[2]{{%
\expandafter\ifx\csname l@#1\endcsname\relax
\typeout{** WARNING: IEEEtran.bst: No hyphenation pattern has been}%
\typeout{** loaded for the language `#1'. Using the pattern for}%
\typeout{** the default language instead.}%
\else
\language=\csname l@#1\endcsname
\fi
#2}}
\providecommand{\BIBdecl}{\relax}
\BIBdecl

\bibitem{GDPR}
``General data protection regulation ({GDPR}),'' \url{https://gdpr.eu/what-is-gdpr}.

\bibitem{Tomashenko_2020}
N.~Tomashenko, X.~Wang, E.~Vincent, J.~Patino, B.~M.~L. Srivastava, P.-G. Noé, A.~Nautsch, N.~Evans, J.~Yamagishi, B.~O’Brien, A.~Chanclu, J.-F. Bonastre, M.~Todisco, and M.~Maouche, ``The voiceprivacy 2020 challenge: Results and findings,'' \emph{Computer Speech \& Language}, vol.~74, p. 101362, Jul. 2022.

\bibitem{tomashenko2022voiceprivacy2022challengeevaluation}
N.~Tomashenko, X.~Miao, P.~Champion, S.~Meyer, X.~Wang, E.~Vincent, M.~Panariello, N.~Evans, J.~Yamagishi, and M.~Todisco, ``The voiceprivacy 2024 challenge evaluation plan,'' \emph{arXiv preprint arXiv:2404.02677}, 2024.

\bibitem{10603395}
M.~Panariello, N.~Tomashenko, X.~Wang, X.~Miao, P.~Champion, H.~Nourtel, M.~Todisco, N.~Evans, E.~Vincent, and J.~Yamagishi, ``The {VoicePrivacy} 2022 challenge: Progress and perspectives in voice anonymisation,'' \emph{IEEE/ACM Transactions on Audio, Speech, and Language Processing}, pp. 1--14, 2024.

\bibitem{tomashenko2024voiceprivacy}
N.~Tomashenko, X.~Miao, P.~Champion, S.~Meyer, X.~Wang, E.~Vincent, M.~Panariello, N.~Evans, J.~Yamagishi, and M.~Todisco, ``The {VoicePrivacy} 2024 challenge evaluation plan,'' \emph{arXiv preprint arXiv:2404.02677}, 2024.

\bibitem{champion2023}
P.~Champion, S.~Ouni, D.~Jouvet, and A.~Larcher, ``Anonymizing speech: Evaluating and designing speaker anonymization techniques,'' \emph{arXiv preprint}, vol. arXiv:2308.04455v4, 2023.

\bibitem{panariello2024speakeranonymizationusingneural}
M.~Panariello, F.~Nespoli, M.~Todisco, and N.~Evans, ``Speaker anonymization using neural audio codec language models,'' 2024.

\bibitem{meyer2023prosody}
S.~Meyer, F.~Lux, J.~Koch, P.~Denisov, P.~Tilli, and N.~T. Vu, ``Prosody is not identity: A speaker anonymization approach using prosody cloning,'' in \emph{Proc. IEEE ICASSP}.\hskip 1em plus 0.5em minus 0.4em\relax IEEE, 2023, pp. 1--5.

\bibitem{shamsabadi2022differentially}
A.~S. Shamsabadi, B.~M.~L. Srivastava, A.~Bellet, N.~Vauquier, E.~Vincent, M.~Maouche, M.~Tommasi, and N.~Papernot, ``{Differentially private speaker anonymization},'' \emph{{Proceedings on Privacy Enhancing Technologies}}, vol. 2023, no.~1, Jan. 2023.

\bibitem{mawalim2022speaker}
C.~O. Mawalim, K.~Galajit, J.~Karnjana, S.~Kidani, and M.~Unoki, ``Speaker anonymization by modifying fundamental frequency and x-vector singular value,'' \emph{Computer Speech \& Language}, vol.~73, p. 101326, 2022.

\bibitem{yao2024musa}
J.~Yao, Q.~Wang, P.~Guo, Z.~Ning, Y.~Yang, Y.~Pan, and L.~Xie, ``{MUSA}: Multi-lingual speaker anonymization via serial disentanglement,'' \emph{arXiv preprint arXiv:2407.11629}, 2024.

\bibitem{miao22_odyssey}
X.~Miao, X.~Wang, E.~Cooper, J.~Yamagishi, and N.~Tomashenko, ``Language-independent speaker anonymization approach using self-supervised pre-trained models,'' in \emph{Proc. The Speaker and Language Recognition Workshop (Odyssey 2022)}, 2022, pp. 279--286.

\bibitem{miao2023language}
------, ``Speaker anonymization using orthogonal householder neural network,'' \emph{IEEE/ACM Trans. Audio, Speech, and Language Processing}, vol.~31, pp. 3681--3695, 2023.

\bibitem{yao2024distinctive}
J.~Yao, Q.~Wang, P.~Guo, Z.~Ning, and L.~Xie, ``Distinctive and natural speaker anonymization via singular value transformation-assisted matrix,'' \emph{IEEE/ACM Transactions on Audio, Speech, and Language Processing}, 2024.

\bibitem{xinyuan2024hltcoe}
H.~L. Xinyuan, Z.~Cai, A.~Garg, K.~Duh, L.~P. Garc{\'\i}a-Perera, S.~Khudanpur, N.~Andrews, and M.~Wiesner, ``Hltcoe jhu submission to the voice privacy challenge 2024,'' \emph{arXiv preprint arXiv:2409.08913}, 2024.

\bibitem{cai2024privacy}
Z.~Cai, H.~L. Xinyuan, A.~Garg, L.~P. Garc{\'\i}a-Perera, K.~Duh, S.~Khudanpur, N.~Andrews, and M.~Wiesner, ``Privacy versus emotion preservation trade-offs in emotion-preserving speaker anonymization,'' in \emph{2024 IEEE Spoken Language Technology Workshop (SLT)}.\hskip 1em plus 0.5em minus 0.4em\relax IEEE, 2024, pp. 409--414.

\bibitem{ghosh2024anonymising}
S.~Ghosh, M.~Jouaiti, A.~Das, Y.~Sinha, T.~Polzehl, I.~Siegert, and S.~Stober, ``Anonymising elderly and pathological speech: Voice conversion using ddsp and query-by-example,'' \emph{Interspeech}, 2024.

\bibitem{zhang2023voicepm}
Z.~Shaohu, L.~Zhouyu, and D.~Anupam, ``{VoicePM}: A robust privacy measurement on voice anonymity,'' in \emph{Proc. 16th ACM Conference on Security and Privacy in Wireless and Mobile Networks (WiSec)}, 2023, p. 215–226.

\bibitem{miao2024benchmark}
X.~Miao, R.~Tao, C.~Zeng, and X.~Wang, ``A benchmark for multi-speaker anonymization,'' \emph{IEEE Transactions on Information Forensics and Security}, vol.~20, pp. 3819--3833, 2025.

\bibitem{miao2025adapting}
X.~Miao, Y.~Zhang, X.~Wang, N.~Tomashenko, D.~C.~L. Soh, and I.~Mcloughlin, ``Adapting general disentanglement-based speaker anonymization for enhanced emotion preservation,'' \emph{Computer Speech \& Language}, p. 101810, 2025.

\bibitem{yao2025easy}
J.~Yao, H.~Liu, E.~S. Chng, and L.~Xie, ``Easy: Emotion-aware speaker anonymization via factorized distillation,'' \emph{Accepted by Interspeech 2025}, 2025.

\bibitem{das2025}
C.~Franzreb, A.~Das, T.~Polzehl, and S.~Möller, ``Private knn-vc: Interpretable anonymization of converted speech,'' in \emph{Accepted by Interspeech 2025}, 2025.

\bibitem{Desplanques_2020}
B.~Desplanques, J.~Thienpondt, and K.~Demuynck, ``Ecapa-tdnn: Emphasized channel attention, propagation and aggregation in tdnn based speaker verification,'' in \emph{Interspeech 2020}.\hskip 1em plus 0.5em minus 0.4em\relax ISCA, 2020.

\bibitem{tomashenko2024voiceprivacyattackerchallengeevaluation}
N.~Tomashenko, X.~Miao, E.~Vincent, and J.~Yamagishi, ``The first voiceprivacy attacker challenge evaluation plan,'' 2024.

\bibitem{Tomashenko_2025}
------, ``The first voiceprivacy attacker challenge,'' in \emph{ICASSP 2025 - 2025 IEEE International Conference on Acoustics, Speech and Signal Processing (ICASSP)}.\hskip 1em plus 0.5em minus 0.4em\relax IEEE, Apr. 2025, p. 1–2.

\bibitem{li2025specwav}
Y.~Li, Y.~Zheng, Z.~Guo, Y.~Wang, J.~Yin, and H.~Fei, ``Specwav-attack: Leveraging spectrogram resizing and wav2vec 2.0 for attacking anonymized speech,'' in \emph{ICASSP 2025-2025 IEEE International Conference on Acoustics, Speech and Signal Processing (ICASSP)}.\hskip 1em plus 0.5em minus 0.4em\relax IEEE, 2025, pp. 1--2.

\bibitem{baevski2020wav2vec}
A.~Baevski, Y.~Zhou, A.~Mohamed, and M.~Auli, ``wav2vec 2.0: A framework for self-supervised learning of speech representations,'' in \emph{Proc. NIPS}, vol.~33, 2020, pp. 12\,449--12\,460.

\bibitem{hu2022lora}
E.~J. Hu, Y.~Shen, P.~Wallis, Z.~Allen-Zhu, Y.~Li, S.~Wang, L.~Wang, W.~Chen \emph{et~al.}, ``Lora: Low-rank adaptation of large language models.'' \emph{ICLR}, vol.~1, no.~2, p.~3, 2022.

\bibitem{he2016deep}
K.~He, X.~Zhang, S.~Ren, and J.~Sun, ``Deep residual learning for image recognition,'' in \emph{Proceedings of the IEEE conference on computer vision and pattern recognition}, 2016, pp. 770--778.

\bibitem{lyu2025fast}
X.~Lyu, Y.~Wang, T.~Zhao, and H.~Liu, ``Fast adaptation of pretrained speaker verification system for source speaker tracking,'' in \emph{ICASSP 2025-2025 IEEE International Conference on Acoustics, Speech and Signal Processing (ICASSP)}.\hskip 1em plus 0.5em minus 0.4em\relax IEEE, 2025, pp. 1--2.

\bibitem{mawalim2025fine}
C.~O. Mawalim, A.~Adila, and M.~Unoki, ``Fine-tuning titanet-large model for speaker anonymization attacker systems,'' in \emph{ICASSP 2025-2025 IEEE International Conference on Acoustics, Speech and Signal Processing (ICASSP)}.\hskip 1em plus 0.5em minus 0.4em\relax IEEE, 2025, pp. 1--2.

\bibitem{zhang2025attacking}
Y.~Zhang, Z.~Bi, F.~Xiao, X.~Yang, Q.~Zhu, and J.~Guan, ``Attacking voice anonymization systems with augmented feature and speaker identity difference,'' in \emph{ICASSP 2025-2025 IEEE International Conference on Acoustics, Speech and Signal Processing (ICASSP)}.\hskip 1em plus 0.5em minus 0.4em\relax IEEE, 2025, pp. 1--2.

\bibitem{prince2007probabilistic}
S.~J. Prince and J.~H. Elder, ``Probabilistic linear discriminant analysis for inferences about identity,'' in \emph{2007 IEEE 11th international conference on computer vision}.\hskip 1em plus 0.5em minus 0.4em\relax IEEE, 2007, pp. 1--8.

\bibitem{ioffe2006probabilistic}
S.~Ioffe, ``Probabilistic linear discriminant analysis,'' in \emph{European Conference on Computer Vision}.\hskip 1em plus 0.5em minus 0.4em\relax Springer, 2006, pp. 531--542.

\bibitem{Park_2019}
D.~S. Park, W.~Chan, Y.~Zhang, C.-C. Chiu, B.~Zoph, E.~D. Cubuk, and Q.~V. Le, ``Specaugment: A simple data augmentation method for automatic speech recognition,'' in \emph{Interspeech 2019}.\hskip 1em plus 0.5em minus 0.4em\relax ISCA, Sep. 2019.

\bibitem{xinyuan2025hltcoe}
H.~L. Xinyuan, A.~Garg, Z.~Cai, K.~Duh, L.~P. Garc{\'\i}a-Perera, S.~Khudanpur, N.~Andrews, and M.~Wiesner, ``Hltcoe submission to the voiceprivacy attacker challenge,'' in \emph{ICASSP 2025-2025 IEEE International Conference on Acoustics, Speech and Signal Processing (ICASSP)}.\hskip 1em plus 0.5em minus 0.4em\relax IEEE, 2025, pp. 1--2.

\bibitem{gaznepoglu2025disentanglement}
U.~E. Gaznepoglu and N.~Peters, ``Why disentanglement-based speaker anonymization systems fail at preserving emotions?'' in \emph{ICASSP 2025 - 2025 IEEE International Conference on Acoustics, Speech and Signal Processing (ICASSP)}, 2025, pp. 1--5.

\bibitem{tomashenko2025analysis}
N.~Tomashenko, E.~Vincent, and M.~Tommasi, ``Analysis of speech temporal dynamics in the context of speaker verification and voice anonymization,'' in \emph{ICASSP 2025-2025 IEEE International Conference on Acoustics, Speech and Signal Processing (ICASSP)}.\hskip 1em plus 0.5em minus 0.4em\relax IEEE, 2025, pp. 1--5.

\bibitem{tomashenko2025exploiting}
------, ``Exploiting context-dependent duration features for voice anonymization attack systems,'' in \emph{Interspeech 2025}, 2025.

\bibitem{bauer2025inferenceattacksxvectorspeaker}
W.~Bao, M.~Jadhav, and V.~Bindschaedler, ``{ Inference Attacks for X-Vector Speaker Anonymization },'' in \emph{2025 IEEE Security and Privacy Workshops (SPW)}, 2025, pp. 152--159.

\bibitem{panayotov2015librispeech}
V.~Panayotov, G.~Chen, D.~Povey, and S.~Khudanpur, ``{LibriSpeech}: an {ASR} corpus based on public domain audio books,'' in \emph{Proc. ICASSP}.\hskip 1em plus 0.5em minus 0.4em\relax IEEE, 2015, pp. 5206--5210.

\bibitem{b3}
S.~Meyer, P.~Tilli, P.~Denisov, F.~Lux, J.~Koch, and N.~T. Vu, ``Anonymizing speech with generative adversarial networks to preserve speaker privacy,'' in \emph{2022 IEEE Spoken Language Technology Workshop (SLT)}, 2023, pp. 912--919.

\bibitem{yao2024npu}
J.~Yao, N.~Kuzmin, Q.~Wang, P.~Guo, Z.~Ning, D.~Guo, K.~A. Lee, E.-S. Chng, and L.~Xie, ``{NPU-NTU System for Voice Privacy 2024 Challenge},'' \emph{arXiv preprint arXiv:2409.04173}, 2024.

\bibitem{Kuzmin2024ntu}
\BIBentryALTinterwordspacing
N.~Kuzmin, H.-T. Luong, J.~Yao, L.~Xie, and K.~A. Lee, ``{NTU-NPU System for Voice Privacy 2024 Challenge},'' \emph{SPSC 2024}, 2024. [Online]. Available: \url{https://www.voiceprivacychallenge.org/vp2024/docs/T12_____NTU-NPU_System_for_Voice_Privacy_2024_Challenge.pdf}
\BIBentrySTDinterwordspacing

\bibitem{Gu2024ustc}
\BIBentryALTinterwordspacing
W.~Gu, Z.~Liu, L.~Chen, R.~Wang, C.~Guo, W.~Guo, K.~A. Lee, and Z.-H. Ling, ``{USTC-PolyU system for the VoicePrivacy 2024 Challenge},'' \emph{SPSC 2024}, 2024. [Online]. Available: \url{https://www.voiceprivacychallenge.org/vp2024/docs/T25_____USTC-PolyU_system_for_the_VoicePrivacy_2024_Challenge.pdf}
\BIBentrySTDinterwordspacing

\bibitem{park2019specaugment}
D.~S. Park, W.~Chan, Y.~Zhang, C.-C. Chiu, B.~Zoph, E.~D. Cubuk, and Q.~V. Le, ``Specaugment: A simple data augmentation method for automatic speech recognition,'' \emph{Interspeech 2019}, Sep 2019.

\bibitem{radford2022robustspeechrecognitionlargescale}
A.~Radford, J.~W. Kim, T.~Xu, G.~Brockman, C.~McLeavey, and I.~Sutskever, ``Robust speech recognition via large-scale weak supervision,'' 2022.

\bibitem{muller2007dynamic}
M.~M{\"u}ller, ``Dynamic time warping,'' \emph{Information retrieval for music and motion}, pp. 69--84, 2007.

\end{thebibliography}

\end{document}